# Phonon Engineering of the Specific Heat of Twisted Bilayer Graphene: The Role of the Out-of-Plane Phonon Modes

## Alexandr I. Cocemasov[1,2], Denis L. Nika[1,2,*] and Alexander A. Balandin[1,♦]

[1]Nano-Device Laboratory and Phonon Optimized Engineered Materials (POEM) Center,
University of California – Riverside, Riverside, California 92521 USA

[2]E. Pokatilov Laboratory of Physics and Engineering of Nanomaterials, Department of
Theoretical Physics, Moldova State University, Chisinau, MD-2009, Republic of Moldova

## Abstract

We investigated theoretically the specific heat of graphene, bilayer graphene and twisted bilayer graphene taking into account the exact phonon dispersion and density of states for each polarization branch. It is shown that *contrary* to a conventional believe the dispersion of the out-of-plane acoustic phonons – referred to as *ZA* phonons – deviates strongly from a parabolic law starting from the frequencies as low as ~100 cm$^{-1}$. This leads to the frequency-dependent *ZA* phonon density of states and the breakdown of the linear dependence of the specific heat on temperature *T*. We established that *ZA* phonons determine the specific heat for $T \leq 200$ K while contributions from both in-plane and out-of-plane *acoustic* phonons are dominant for $200$ K $\leq T \leq 500$ K. In the high-temperature limit, *T*>1000 K, the optical and acoustic phonons contribute approximately equally to the specific heat. The Debye temperature for graphene and twisted bilayer graphene was calculated to be around ~1861 – 1864 K. Our results suggest that the thermodynamic properties of materials such as bilayer graphene can be controlled at the atomic scale by rotation of the sp$^2$-carbon planes.

**KEYWORDS:** phonon density of states, graphene, bilayer, twisted graphene, heat capacity

---
[*] Corresponding authors: (DLN) dnica@ee.ucr.edu and (AAB) balandin@ee.ucr.edu

Alexandr I. Cocemasov, Denis L. Nika and Alexander A. Balandin (UC Riverside, 2015)

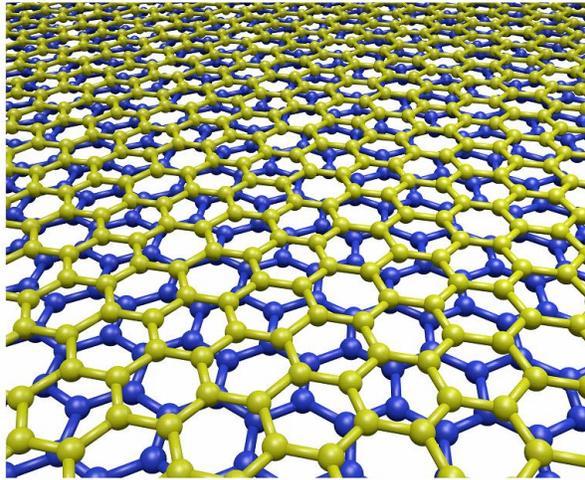
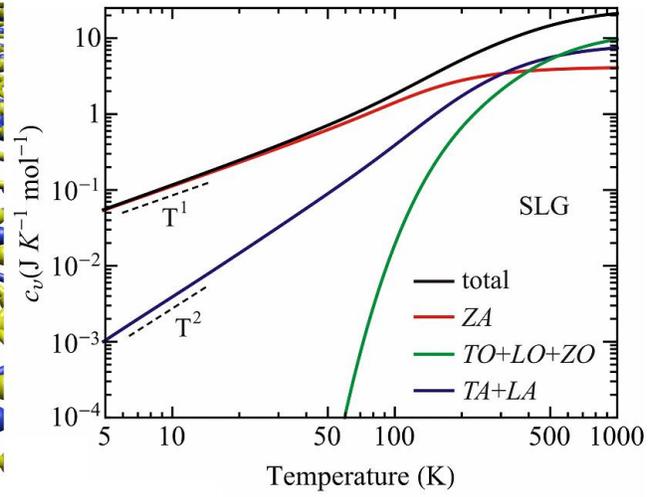

**Contents Image**

**INTRODUCTION**

The discovery of the unusually high thermal conductivity, $K$, of suspended graphene and few-layer graphene (FLG) [1-4] resulted in a surge of interest to thermal properties of two-dimensional (2-D) materials [5-15]. The optothermal studies conducted with the help of micro-Raman spectroscopy revealed the thermal conductivity $K \sim 2000 - 5000$ W/mK (depending on the sample size and quality), near room temperature (RT) for the large graphene samples suspended across trenches in $SiO_2$ wafers [1-4] or on the transmission electron microscopy (TEM) grids [5, 8]. These values are above the bulk graphite limit of $K=2000$ W/mK for basal planes at RT [1]. A more recent study using a different experimental technique – electrical thermal bridge – of the residue-free suspended graphene also obtained the thermal conductivity above the bulk graphite limit ($K \sim 2430$ W/mK at $T=335$ K) [15]. Graphene sample supported on substrates have lower thermal conductivity, e.g. $K \sim 600$ W/mK for graphene on $Si/SiO_2$ at $T=300$ K [6]. However, even on a substrate or embedded in matrix, FLG reveals higher $K$ as compared to thin films of comparable thickness made of other materials [1, 16].

There is an inherent ambiguity in the definition of the thermal conductivity of graphene related in the thickness value of the atomic plane $h$ (most of studies use $h=0.35$ nm, which originates from the carbon-carbon bond length). Despite this ambiguity and unavoidable experimental data scatter due to different sizes and quality of the samples, there is a growing consensus among theorists that the phonon thermal properties of graphene can be fundamentally different from those of three-dimensional (3-D) bulk crystals [1, 4, 9-11, 14, 17 - 22]. The latter can be attributed to the 2-D nature of the phonon density of states in graphene and resulting exceptionally long phonon mean free path (MFP) for the long wavelength phonons. Recent theoretical studies suggested that graphene samples with the 100-μm length [19] or even 1-mm length [20] are required in order to recover the intrinsic thermal conductivity of graphene. In both reports the $K$ values were substantially larger than the bulk graphite limit ($K=5800$ W/mK at $T=300$ K in Ref. [19]). The specifics of the phonon dispersion and relative contributions of different phonon polarization are important for gaining complete understanding of thermal properties of 2-D materials.

In this paper, we report the results of our investigation of the specific heat, $c_v$, of single layer graphene (SLG), bilayer graphene (BLG) and twisted bilayer graphene (T-BLG). This thermodynamic characteristic can be determined rather accurately and it suffers less from the ambiguity of the phonon transport characteristics. The focus of our present study is on elucidating the role of the out-of-plane phonons in determining the specific heat in different temperature ranges and revealing the influences of the 2-D phonon density of states (PDOS). In addition to a numerical solution we also provide a simple analytical formula for calculating $c_v(T)$ for graphene, BLG and T-BLG with parameters extracted from the Born-von Karman model of the lattice vibrations. We have earlier proposed a possibility of engineering phonon dispersion and materials properties by twisting of the atomic planes in T-BLG [23-24]. The first experimental studies of heat conduction in suspended T-BLG confirmed that twisting substantially reduces $K$ owing to the increased scattering phase space available for phonons in T-BLG as compared to the Bernal-stacked BLG [25]. In this work the approach for controlling phonon properties at atomic level by rotating $sp^2$ carbon planes is treated in a broader context. The knowledge of the specific heat and Debye temperature of SLG, BLG and T-BLG is important for practical applications of these materials as fillers in thermal pasts [26-29] and thermal graphene laminate coatings [16, 30-33].

## 1. PHONON DENSITY OF STATES IN SLG, BLG and T-BLG

It is known that SLG reveals four in-plane phonon branches: transverse/longitudinal acoustic (*TA/LA*) and optic (*TO/LO*) branches with the atomic displacements in the graphene plane, and two out-of-plane acoustic (*ZA*) and optic (*ZO*) branches with the displacements perpendicular to the graphene plane. The in-plane acoustic branches are characterized by the linear energy dispersions over the most part of the Brillion zone (BZ) except near the zone edge while the out-of-plane *ZA* branch demonstrates a quadratic dispersion near the zone center $q = 0$, where $q$ is the phonon wavenumber. The number of the phonon branches in BLG is doubled: six additional branches possess non-zero frequency at $q = 0$ and at the low frequencies are affected by the inter-layer interactions [3, 10, 22, 23]. The emergence of many folded hybrid phonon branches in T-BLG was explained by the change of the unit cell size and a corresponding modification of the reciprocal space geometry. The number of the polarization branches and their dispersion in T-BLG depend strongly on the rotation angle [23].

We determine PDOS in SLG, BLG and T-BLG using the phonon dispersions calculated in the framework of the lattice dynamics theory. For the intralayer carbon-carbon interaction we use the Born – von Karman (BvK) force constant approach [23]. For the interlayer interaction we use the spherically symmetric interatomic potential with the following components of the force constant matrices: $\Phi_{\alpha\beta}^{ij} = -\delta(r^{ij})r_{\alpha}^{ij}r_{\beta}^{ij}/(r^{ij})^2$, where $\delta$ is the force constant of the interlayer coupling, $\vec{r}^{ij}$ is the vector connecting a pair of interacting atoms $(i,j)$; $\alpha$ and $\beta$ designate the components of the Cartesian coordinates. Since the van der Waals coupling strength between the graphene layers is very weak we model the dependence of the force constant $\delta$ on $r^{ij}$ as: $\delta(r^{ij}) = A\exp(-r^{ij}/B)$ with two fitting parameters $A$=573.76 N/m and $B$=0.05 nm [24], determined from comparison with the experimental phonon frequencies from $\Gamma$-A direction in graphite [34].

Twisted bilayer graphene can be generated from BLG by rotating, i.e. twisting, one layer relative to another by an angle $\theta$ in the graphene plane (see Fig. 1(a)). A set of T-BLGs with different rotational angles and commensurate crystal lattice, i.e. lattice possessing a translational symmetry, are determined by the following relation [35]: $\cos\theta(p,n) = (3p^2 + 3pn + n^2/2)/(3p^2 + 3pn + n^2)$, where $p$ and $n$ are coprime positive integer numbers. The number of atoms in a unit cell of T-BLG depends on $\theta$ and it is given by the relation $N = 4(3p^2 + 3pn + n^2)$, if $n$ is not divisible by 3 [23, 24]. For example, the unit cell of the T-BLG with $\theta(2,1) = 13.2°$ contains 76 atoms and its BZ is 19 times smaller than that of BLG without a rotation (see Fig. 1(b) for BZ schematics). A detailed description of the theoretical approach for calculating the phonon modes in SLG, BLG and T-BLG was reported by us recently elsewhere [23, 24]. Here we use this approach for the investigation of the polarization branch dependent PDOS.

The 2D phonon density of states per unit surface area for SLG and FLG is given by:

$$g(\omega) = \sum_s g_s(\omega); \quad g_s(\omega) = \frac{1}{4\pi^2} \sum_{q_x(s,\omega)} \sum_{q_y(s,\omega,q_x)} \frac{\Delta q_x}{\left| \upsilon_y(q_x, q_y, s) \right|}. \quad (1)$$

Here $\omega$ is the phonon frequency, $s$ numerates phonon branches (polarizations), $g_s(\omega)$ is the polarization-dependent phonon density of states, $q_x$ and $q_y$ are components of the 2D phonon wave vector, $\upsilon_y = \partial\omega / \partial q_y$ is the $y$-component of the phonon group velocity, $\Delta q_x$ is the interval between two neighboring $q_x$ points. In order to determine $g(\omega)$ from Eq. (1) we calculated $\omega_s(q_x, q_y)$ in 40 000 points $(q_x, q_y)$ (200 × 200 grid) uniformly distributed over a 1/4$^{th}$ part of the BZ, shown as green segment in Fig. 1 for T-BLG with a 13.2° rotation. We checked that increasing the number of points by a factor of 4 does not change the numerical results. Our analysis also showed that 22500 BZ points (150 × 150 grid) is already sufficient for convergence and obtaining accurate results for PDOS.

The PDOS for *LA*, *TA* and *ZA* branches in graphene can be derived analytically in the isotropic approximation of the linear frequency dispersion for *LA* and *TA* branches $\omega_{LA,TA}(q) = \upsilon_{LA,TA} \times q$ and quadratic dispersion for *ZA* branch $\omega_{ZA}(q) = \alpha \times q^2$ in the entire BZ, where $\upsilon_{LA}$ ($\upsilon_{TA}$) is the $q$-independent group velocity and $\alpha$ is a parameter. In this approximation the PDOS per unit surface area takes the form:

$$g_{LA,TA}^{isot}(\omega) = \frac{\omega}{2\pi\upsilon_{LA,TA}^2}, \; g_{ZA}^{isot}(\omega) = \frac{1}{4\pi\alpha}. \quad (3)$$

The phonon energy dispersion along *Γ-M* direction of BZ are shown in Fig. 2 for SLG (panel a), BLG (panel b) and T-BLG with 13.2$^0$ rotation (panel c). The red triangles indicate the experimental phonon frequencies of graphite, reproduced from Ref. [36]. One can conclude that our lattice dynamics model provides a good agreement between the theoretical and experimental phonon frequencies. The dispersion of *ZA* branch in SLG and BLG can be divided into two distinctive regions: (I) region with the quadratic dispersion $q<5.2$ nm$^{-1}$ and (II) region with almost linear dispersion 5.2 nm$^{-1}$ $<q<13.0$ nm$^{-1}$. Therefore we can improve Eq. (3) for *ZA* PDOS accounting for both quadratic and linear dispersion regions:

$$g_{ZA}^{isot}(\omega) = \frac{1}{4\pi\alpha}\Theta(\omega - \omega_c) + \frac{\omega}{2\pi\upsilon_{ZA}^2}\Theta(\omega_c - \omega), \quad (4)$$

where $\omega_c$ is the phonon frequency separating region (I) from region (II), $\Theta$ is the Heaviside's step function and $\upsilon_{ZA}$ is the velocity of *ZA* modes in region II.

In Fig. 3 we analyze partial contribution of *LA* (green), *TA* (blue), *ZA* (red) and *ZO* (magenta) branches to the total PDOS (solid black) for SLG. There are 7 pronounced peaks in the PDOS curve. The peaks at 452, 605 and 638 cm$^{-1}$ are associated with *ZA*, *TA* and *ZO* phonons at BZ edge, correspondingly. The peak frequency of *LA* branch ~ 1192 cm$^{-1}$ is smaller than *LA* phonon frequency at *M* point ~ 1287 cm$^{-1}$ (see Fig. 2 (a)). The main contributors to this peak are the low-velocity phonons from the directions near BZ edge. The PDOS peak at 889 cm$^{-1}$ is related to *ZO* phonon at BZ center (*Γ*-point), while *TO* and *LO* phonons at BZ center and BZ edge are responsible for peaks at 1350 and 1585 cm$^{-1}$. In the case of AB-stacked BLG all phonon branches except *ZA* become nearly doubly degenerated over entire BZ and intensity of PDOS is by a factor of ~ 2 larger than in SLG (see dashed black line in Fig. 3). The new peak at 91 cm$^{-1}$ also appears in AB-BLG. This peak is associated with $ZA_2$ phonon at *Γ*-point (see Fig. 2(b)). In T-BLG the frequency of the $ZA_2$ phonon at *Γ*-point depends on the angle of rotation due to the changes in the interlayer coupling [23], while the overall $g(\omega)$ remains the same as for AB-BLG with slightly shifted $ZA_2$ peak. For example, the $ZA_2$ peak shifts to 89.3 cm$^{-1}$ for T-BLG with 13.2$^0$ rotation and to 89.5 cm$^{-1}$ for T-BLG with 21.8$^0$ rotation.

In Fig. 4(a) we compare accurate *LA*, *TA* and *ZA* PDOS in SLG calculated from Eq. (1) (solid lines) with those obtained in the isotropic approximation using Eqs. (3-4) (dashed lines). The parameters of the isotropic approximation were extracted from the actual phonon dispersions: $\upsilon_{TA}$ =13.5 km/s, $\upsilon_{LA}$ =20.4 km/s, $\upsilon_{ZA}$ =7.9 km/s , $\alpha$ =0.62x10$^{-6}$ m$^2$/s and $\omega_c$ = 90 cm$^{-1}$. At small energies both sets of the curves almost coincide. The rise of the phonon energy leads to the increasing difference between the curves due to deviation of *LA/TA* dispersion from the linear law and *ZA* dispersion from the quadratic law. It is evident from Fig. 4(a), that the high PDOS peaks cannot be described by the isotropic analytical expressions of Eqs. (3-4), and the deviation from the accurate PDOS curves become significant for the frequencies above ~250 cm$^{-1}$ for *ZA*, ~300 cm$^{-1}$ for *TA* and ~600 cm$^{-1}$ for *LA* polarizations. The strong influence of non-parabolicity of *ZA* branch on *ZA* PDOS is illustrated in Fig. 4(b). We show *ZA* PDOS as a function of the phonon frequency for two cases: (1) a small segment near BZ center is described by the isotropic and parabolic dispersions while in the rest of BZ the PDOS is calculated using the actual phonon dispersion obtained from BvK model of lattice vibrations (anisotropic and non-parabolic *ZA* dispersion), and (2) a small segment near BZ center is anisotropic while the rest is isotropic, which is opposite to case (1).

The difference between two cases is only in the phonon dispersions used. Both $g(\omega)$ curves were calculated from Eq. (1). In the frequency range where the dispersion is parabolic, $g(\omega)$ should be constant and equal to $1/(4\pi\alpha)$ (see Eq. 3). The maximum *ZA* phonon frequency in the small segment near BZ center, shown as white tetragon in Fig. 4, is ~100 cm$^{-1}$ and is almost the same for the parabolic and BvK dispersions. Both $g(\omega)$ curves (red and blue) are practically independent on $\omega$ up to ~90 cm$^{-1}$ and coincide with the dashed line $\omega = 1/(4\pi\alpha)$. Therefore, one can conclude that in the region of the phonon wave vectors not far from the BZ center ($\omega < 90$ cm$^{-1}$) the *ZA* BvK phonon dispersions are almost parabolic and isotropic. For the higher phonon frequencies the difference between two curves reinforces due to the strong anisotropy and non-parabolicity of *ZA* branches.

The Debye temperature is one of the most important parameters describing the thermal properties of solids. We can estimate the Debye temperature in SLG, BLG and T-BLG, using Eqs. (1, 3, 4). In the Debye model the total number of phonon states $N_S$ is given by:

$$N_S = A \sum_{s=LA,TA,ZA} \int_0^{\omega_D} g_s^{isot}(\omega)d\omega = A(R_1\omega_D^2 + R_2), \qquad (5)$$

where $A$ is the surface area, $R_1 = \sum_{s=LA,TA,ZA} 1/(4\pi\upsilon_s^2)$ and $R_2 = \frac{\omega_c}{4\pi\alpha}(1 - \frac{\alpha\omega_c}{\upsilon_{ZA}^2})$. Summation in Eq. (5) is performed over all acoustic branches. The contribution from the optical phonons is assumed to be zero in the Debye model. Using real PDOS, described by Eq. (1), $N_S$ takes the form:

$$N_S = A\sum_s \int_{\omega_{s,min}}^{\omega_{s,max}} g_s(\omega)d\omega. \qquad (6)$$

Here $s$ enumerates both the acoustic and optical phonon branches. From Eqs. (5-6) one can calculate the Debye's frequency as:

$$\omega_D = \sqrt{\frac{\sum\limits_s \int\limits_{\omega_{s,min}}^{\omega_{s,max}} g_s(\omega)d\omega - R_2}{R_1}}. \qquad (7)$$

From Eq. (7) we estimated the Debye frequency in SLG, BLG and T-BLG. The Debye frequency is extremely high in all cases and differs rather weakly in SLG, BLG and T-BLG: $\omega_D(\text{SLG}) = 1294.1$ cm$^{-1}$, $\omega_D(\text{BLG}) = 1293.7$ cm$^{-1}$ and $\omega_D(\text{T-BLG}) = 1295.7$ cm$^{-1}$, respectively. The weak dependence of $\omega_D$ on the number of atomic layers and twisting is an expected result since the main difference between phonon properties of SLG, BLG and T-BLG are due to *ZA* modes with frequencies $\omega << \omega_D$. These modes are completely populated at temperatures much lower than the Debye temperature $T_D = \hbar\omega_D / k_B$, where $k_B$ and $\hbar$ is the Boltzmann's and reduced Planck's constants, respectively: $T_D(\text{SLG}) = 1862$ K, $T_D(\text{BLG}) = 1861$ K and $T_D(\text{T-BLG}) = 1864$ K. These values of $T_D$ exceed those for the most of materials and are only slightly smaller than $T_D$ of diamond ~ 2000 K [37].

For comparison, in Table I we provide values of $T_D$ for graphite and graphene available in the literature. The separate in-plane and out-of-plane Debye temperatures were calculated from the Debye frequencies using Eq. (7) by summation over the in-plane or out-of-plane branches, respectively. The obtained values are in the same range as those calculated for graphite and graphene in Refs. [41, 44] using the lattice dynamics [41] and Green's functions theory [44]. At the same time, the smaller values of $T_D = 1495$ K and 1045 K were estimated experimentally in Refs [42-43]. We attribute the discrepancy between the theoretical and experimental data to the variations in the contribution of the out-of-plane phonons due to the specific conditions of the experiments [42-43]. For example, authors of Ref. [43] assumed that the weak interlayer bonds between graphene and ruthenium substrate modes effectively scatter impinging He atomic beam when collecting diffraction spectra.

## 2. SPECIFIC HEAT OF SLG, BLG AND T-BLG: IMPACT OF PDOS

The phonon density of states is a key parameter determining the phonon – assisted processes in graphene and related materials. Knowing the frequency distribution of the polarization-specific PDOS we can address particularly interesting questions: (1) how the phonons of different polarizations determine the functional dependence of phonon specific heat on the temperature and (2) how this dependence changes while going from SLG to BLG and from BLG to T-BLG? The phonon specific heat at a constant volume $c_V$ is given by the following expression [45, 46]:

$$c_V(T) = \sum_s c_{s,V}(T); \ c_{s,V}(T) = \frac{3N_A}{k_B T^2} \int_{\omega_{s,\min}}^{\omega_{s,\max}} \frac{\exp(\frac{\hbar\omega}{k_B T})}{[\exp(\frac{\hbar\omega}{k_B T}) - 1]^2} [\hbar\omega]^2 f_s(\omega) d\omega, \qquad (8)$$

where $\omega_{s,\min} (\omega_{s,\max})$ is the minimum (maximum) phonon frequency for the *s-th* branch, *s* enumerates phonon branches, $T$ is the temperature, $N_A$ is the Avogadro constant, $f_s(\omega) = g_s(\omega) / \sum_j \int_{\omega_{j,\min}}^{\omega_{j,\max}} g_j(\omega) d\omega$ is the two-dimensional normalized PDOS and $c_{s,V}$ is the contribution to total specific heat from *s*-th branch. Analyzing Eq. (8) one can conclude that in the isotropic case of parabolic *ZA* dispersion $\omega_{ZA} \sim q^2$ and linear *LA/TA* dispersions $\omega_{LA,TA} \sim q$, leading to $g_{ZA}(\omega) = \text{const}$ and $g_{LA,TA}(\omega) \sim \omega$, the low-temperature specific heat $c_{ZA,V}$ demonstrates linear dependence on $T$, while $c_{LA,V} (c_{TA,V})$ scales with temperature as $T^2$. At the very low temperatures, *ZA* phonons are mostly populated and $c_V$ should be proportional to $T$. Such temperature dependence of specific heat was reported in Ref. [47] for *T*<100 K. Alofi and Srivastava [48] have theoretically shown that a slight deviation from the linear $T$ dependence occurs due to *LA* and *TA* phonon contribution, and $c_V \sim T^{1.1}$ up to 100 K. However, we have recently established [24] that the anisotropy in the phonon dispersions significantly influences the temperature dependence of specific heat in SLG: $c_V \sim T$ for $T \leq 15\,\text{K}$; $c_V \sim T^{1.1}$ for $15 < T \leq 35\,\text{K}$; $c_V \sim T^{1.3}$ for $15\,\text{K} < T \leq 35\,\text{K}$ and $c_V \sim T^{1.6}$ for $75 < T \leq 240\,\text{K}$. For BLG and T-BLG with 21.8° rotation the dependences $c_V \sim T^{1.3}$ and $c_V \sim T^{1.6}$ were revealed correspondingly for *T*<15 K. These results require a detailed analysis of the interplay between the accurate phonon energy spectra, PDOS and specific heat, which has not been performed to date.

In Fig. 5(a) the temperature dependences of the phonon specific heat $c_{s,V}$ in SLG are shown for different phonon branches: $ZA$ (red), $TA+LA$ (blue) and $ZO+TO+LO$ (green). The contribution of $ZA$ phonons to $c_V$ is dominant up to $T \sim 200$ K. Nevertheless, both $c_{ZA,V}$ and $c_V$ demonstrate deviation from the linear $T$-dependence beginning from $T \sim 15$ K. This is a clear manifestation of the anisotropy and non-parabolicity of $ZA$ dispersions. The power index $n$ of $T^n$-dependence of the specific heat increases faster for total $c_V$ than for $c_{ZA,V}$ due to the contributions from $LA$ and $TA$ phonons revealing $c_{LA,V}$ ($c_{TA,V}$)$\sim T^2$ dependence for $T<100$ K. The contribution of the in-plane phonons to the total $c_V$ increases with the temperature and becomes comparable with that of $ZA$ phonon contribution for $T \sim 250 - 300$ K. This result differs substantially from what is predicted in the framework of the semi-continuum theory for phonon dispersion where the ratio $c_{ZA,V}/$ ($c_{LA,V}+c_{TA,V}$) $\sim 2$ was reported for $T$=300 K [48], which was obtained assuming the constant PDOS for $ZA$ phonons and linear PDOS for $LA$ and $TA$ phonons. The later illustrates that the simplified isotropic models for PDOS in graphene do not capture all the characteristics of the specific heat and thermal conductivity. For temperatures $T \geq 300$ K the contribution of the in-plane acoustic phonons to $c_V$ is larger than that of $ZA$ phonons. The ratio between their contributions ($c_{LA,V}+c_{TA,V}$)$/c_{ZA,V}$ increases from $\sim 1$ at $T$=300 K to $\sim 1.8$ at $T$=1000 K. The contribution from optic phonons $c_{op,V} = c_{LO,V} + c_{TO,V} + c_{ZO,V}$ is very small (<10%) up to $T \sim 180$ K due to the low population of these modes. However, their contribution to $c_V$ grows fast with the temperature: at RT it constitutes $\sim 23\%$, at $T = 500$ K is $\sim 36\%$, while at $T = 1000$ K is $\sim 46\%$.

In Fig. 5(b) we illustrate the contribution from the out-of-plane (red lines) and in-plane (blue lines) phonon modes to the total phonon specific heat of SLG (solid lines) and T-BLG with 13.2° rotation (dashed lines) as a function of temperature. The black lines denote the temperature dependence of the total phonon heat capacity in SLG and T-BLG. The specific heat of the out-of-plane phonons varies with temperature as $T^p$, where $p$=1 for SLG and $p$=1.3 for T-BLG at $T<15$ K. The difference in $p$ is explained by the specifics of the folded phonons in T-BLG (see Fig. 2(c)). In general, the dispersion relations of the "additional" acoustic phonons (denoted as $ZA_2$ in case of non-twisted BLG (see Fig. 2(b)) cannot be described by a parabolic law, even at the phonon wave vectors near BZ center. As a result both $p$(T-BLG) = 1.3 and $p$(AB-BLG) = 1.6 (not shown in Fig. 5) are larger than $p$(SLG). The contribution from the in-plane polarizations to $c_V$ also differs in SLG and T-BLG. At low $T$ in SLG $c_V(T) \sim T^2$, while in T-BLG $c_V(T) \sim T^{2.3}$. This deviation is also due to the appearance of the additional phonon branches in T-BLG with the dispersion relations different from those in SLG.

The results for T-BLG with the rotational angles other than 13.2° (not shown in Fig. 5) are quite similar, because the difference in the absolute values of the total phonon heat capacity for different BLG configurations at temperatures above 5 K is less than 5% and decreases fast with increasing temperature [24]. The low-temperature specific heat of graphite follows the cubic law $c_V(T) \sim T^3$ due to the three-dimensional density of states, thus increasing the number of graphene layers should increase the power factor $n$ in $c_V(T) \sim T^n$ dependence.

The accurate dependence of the specific heat of SLG, BLG and T-BLG on temperature can be adequately approximated by a parabolic function $c_V(T) = aT + bT^2$, where $a$ and $b$ are constants. The extracted values of these constants for two regions of the temperature: $T < 150$ K and $200\,K \leq T \leq 350\,K$ are presented in Table II. In the case of BLG and T-BLG the values of parameters $a$ and $b$ are close $a$(BLG) $\approx a$(T-BLG), $b$(BLG) $\approx b$(T-BLG) and strongly differ from those of SLG. At low temperatures $T < 150$ K, the ratio $b/a$ in BLG/T-BLG is by a factor of ~6.5 larger than that in SLG, indicating a stronger deviation of BLG/T-BLG $ZA$ dispersions from the parabolic law. The difference between the $b/a$ ratios in SLG and BLG/T-BLG practically disappears at higher temperatures of $200\,K \leq T \leq 350\,K$: $b/a$ (SLG) ~ 0.0018 and $b/a$(BLG/T-BLG) ~ 0.0019, where the relative contribution of $ZA$ modes to the specific heat decreases.

### 3.  DISCUSSION: CONTROLLING PHONONS AT ATOMIC SCALE

The obtained results show that $ZA$ phonons dominate the specific heat for $T \leq 200$ K while their contribution become comparable with that from $LA$ and $TA$ phonons in the temperature range 200 K $\leq T \leq$ 500 K. In this sense, the out-of-plane vibrations, which resemble standing waves, are efficient in storing thermal energy. However, this does not automatically mean that *ZA* phonons make the dominant contribution to the thermal conductivity. In the kinetic theory, the phonon thermal conductivity can be written as $K \sim c_V V \Lambda$, where $V$ is the phonon group velocity, $\Lambda = V \tau$ is the phonon MFP and $\tau$ is the combined relaxation time of the phonons. The thermal conductivity, particularly near RT, is affected strongly by the phonon group velocity (defined by the slopes of the dispersion branches) and phonon scattering due to inharmonicity of the crystal lattice and defects. The question of the relative contribution of *ZA*, *LA* and *TA* phonons in different temperature ranges and under different conditions (e.g. supported graphene vs. suspended), is a subject of interesting theoretical debates [4, 10-11, 21, 49-52]. No conclusive experimental evidence has so far been provided.

We have earlier proposed a possibility of controlling heat flow by engineering phonon dispersion in T-BLG [23,53]. The initial experimental studies of thermal conductivity suspended T-BLG performed using the optothermal technique confirmed that twisting substantially reduces $K$ as compared to BLG [25]. The fundamental difference of phonon engineering by twisting from earlier approaches that involved phonon confinement in acoustically mismatched nanostructures, e.g. conventional quantum wells and nanowires [53-56], is a possibility of controlling not only acoustic phonons but optical phonons as well. The twisting of atomic planes results in breaking the unit cells and reducing BZ, which results in the phonon dispersion modification all the way to the optical branch (see Fig. 2(c)). Our present results indicate that SLG, BLG and T-BLG have distinguishably different $c_V$ temperature dependence that can be traced to the PDOS of individual polarization branches. These dependences, which we provided in the analytical form as well, may help in future experimental studies. The possibility of engineering of the acoustic and optical phonon dispersion can be useful for energy storage and thermal management applications [29].

## 4. CONCLUSIONS

We investigated theoretically the phonon density of states for different phonon branches in single-layer, bilayer and twisted bilayer graphene in the framework of the Born-von Karman model. The density of states for *LA*, *TA* and *ZA* phonons have been compared with those obtained in the simplified isotropic model with the linear dispersion for *LA* and *TA* branches and quadratic dispersion for *ZA* branch. Our results show that the isotropic model describes well only the low-frequency part of PDOS with $\omega < 250$ cm$^{-1}$ for *ZA*, $\omega < 300$ cm$^{-1}$ for *TA* and $\omega < 600$ cm$^{-1}$ for *LA* modes. The deviation of the out-of-plane acoustic phonon dispersions from non-parabolic breaks the linear dependence of the specific heat on temperature: in SLG $c_V(SLG) \sim T$ only for *T*<15 K, while specific heat of BLG and T-BLG demonstrates $T^n$ dependence with *n*>1 even at small temperatures ~ 1 K. The partial contribution of the different phonon branches to the specific heat is a function of the temperature: at *T*<200 K the main contributors are *ZA* phonons; in the range 200 K – 500 K, specific heat is determined by *LA*, *TA* and *ZA* phonons, while at *T*>500 K the contribution of optic phonons exceeds 35%. We have found that *T*-dependence of the heat capacity in SLG, BLG and T-BLG can be approximated by a function $aT + bT^2$ at *T*<350 K and have determined parameters *a* and *b* by fitting the accurate $c_V(T)$ curves. The presented results confirm that the accurate phonon density of states is required for both qualitative and quantitative description of the specific heat of SLG, BLG and T-BLG. The dominance of *ZA* phonons in determine the specific heat for *T*≤ 200 K does not imply their leading role in heat conduction, which depends on the phonon mean free path as well. Our results indicate the thermodynamic properties of materials can be controlled at the atomic scale by rotation of the sp$^2$-carbon planes.

**ACKNOWLEDGMENTS**

This work was supported as part of the Spins and Heat in Nanoscale Electronic Systems (SHINES), an Energy Frontier Research Center funded by the U.S. Department of Energy, Office of Science, Basic Energy Sciences (BES) under Award # SC0012670.

**FIGURE CAPTIONS**

Figure 1: (a) Twisted bilayer graphene schematics. (b) Brillouin zones of SLG (blue or yellow hexagon) and T-BLG with 13.2° rotation (green hexagon). $\Gamma$, $M$ and $K$ denote high-symmetry points of T-BLG BZ.

Figure 2: Phonon dispersion along the $\Gamma$-$M$ crystallographic direction in: (a) single-layer graphene, (b) AB-stacked bilayer graphene and (c) twisted bilayer graphene with 13.2° rotation. The red triangles denote experimental phonon frequencies of graphite from Ref. [26].

Figure 3: (a) Total phonon density of states in SLG (black) and contributions from $ZA$ (red), $TA$ (blue), $ZO$ (magenta) and $LA$ (green) phonon branches. (b) $LA$, $TA$ and $ZA$ PDOS calculated from Eq. (1), using actual phonon dispersions (solid lines) and obtained in the isotropic model from Eqs. (3-4) (dashed lines).

Figure 4: $ZA$ phonon density of states in SLG as a function of phonon frequency calculated using different sets of phonon dispersions.

Figure 5: Phonon branch dependent heat capacity as a function of temperature in SLG (a) and T-BLG with 13.2° rotation (b). In the panel (a) the contributions from different branches are denoted as follows: $ZA$ (red), $TO$+$LO$+$ZO$ (green), $TA$+$LA$ (blue). In the panel (b) the red and blue lines denote contributions from the out-of-plane and in-plane phonons, respectively for SLG (solid curves) and $13.2^0$ T-BLG (dashed curves). In both panels the black curves correspond to the total phonon heat capacity.

**Table I**. Debye's temperatures in graphene, few-layer graphene and graphite

| | $T_D$ (K) | Comment | Reference |
|---|---|---|---|
| **graphite** | 1860 | theory: projector augmented wave method + local density approximation | 38 |
| | 2300 (in-plane) | theory: from fitting the thermal expansion coefficient to experimental data | 39 |
| | 800 (out-of-plane) | | |
| | 2300 (in-plane) | theory: from fitting the thermal expansion coefficient to experimental data | 40 |
| | 800 (our-of-plane) | | |
| | 2500 (in-plane) | theory: lattice dynamics; from fitting the specific heat to experimental data | 41 |
| | 950 (out-of-plane) | | |
| **graphene** | 1495 | experiment: suspended and supported | 42 |
| | 1045 | experiment: supported | 43 |
| | 2300 (in-plane) | theory: green functions | 44 |
| | 1287 (out-of-plane) | | |
| | 1862 | theory: BvK lattice dynamics; from PDOS comparison | this work |
| | 2669 (in-plane) | | |
| | 1292 (out-of-plane) | | |
| **BLG** | 1861 | theory: BvK lattice dynamics; from PDOS comparison | this work |
| | 2675 (in-plane) | | |
| | 1295 (out-of-plane) | | |
| **T-BLG ($13.2^0$)** | 1864 | theory: BvK lattice dynamics; from PDOS comparison | this work |
| | 2671 (in-plane) | | |
| | 1293 (out-of-plane) | | |

**Table II**: Constants of parabolic functions $aT + bT^2$ approximating specific heat in SLG, BLG and $13.2^0$ T-BLG.

| | $a$ ($\times 10^{-3}$ $K^{-1}$) | $b$ ($\times 10^{-5}$ $K^{-1}$) |
|---|---|---|
| **$T < 150$ K** | | |
| **SLG** | 11.87 | 7.21 |
| **BLG** | 3.641 | 13.85 |
| **$13.2^0$ T-BLG** | 3.494 | 13.84 |
| **$200$ K $\leq T \leq 350$ K** | | |
| **SLG** | 19.12 | 3.42 |
| **BLG** | 18.56 | 3.57 |
| **$13.2^0$ T-BLG** | 18.36 | 3.6 |

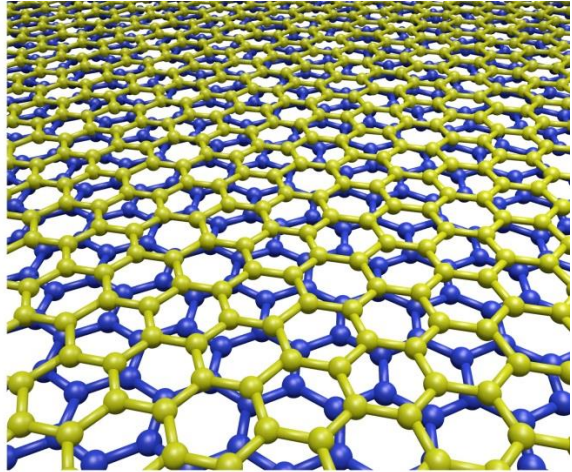

(a)

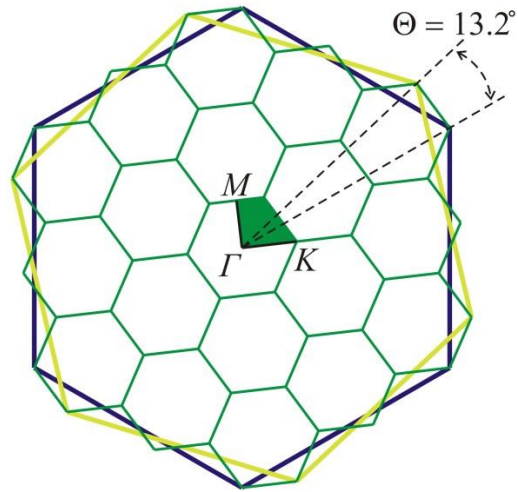

(b)

Figure 1 of 5: A.I. Cocemasov et al.

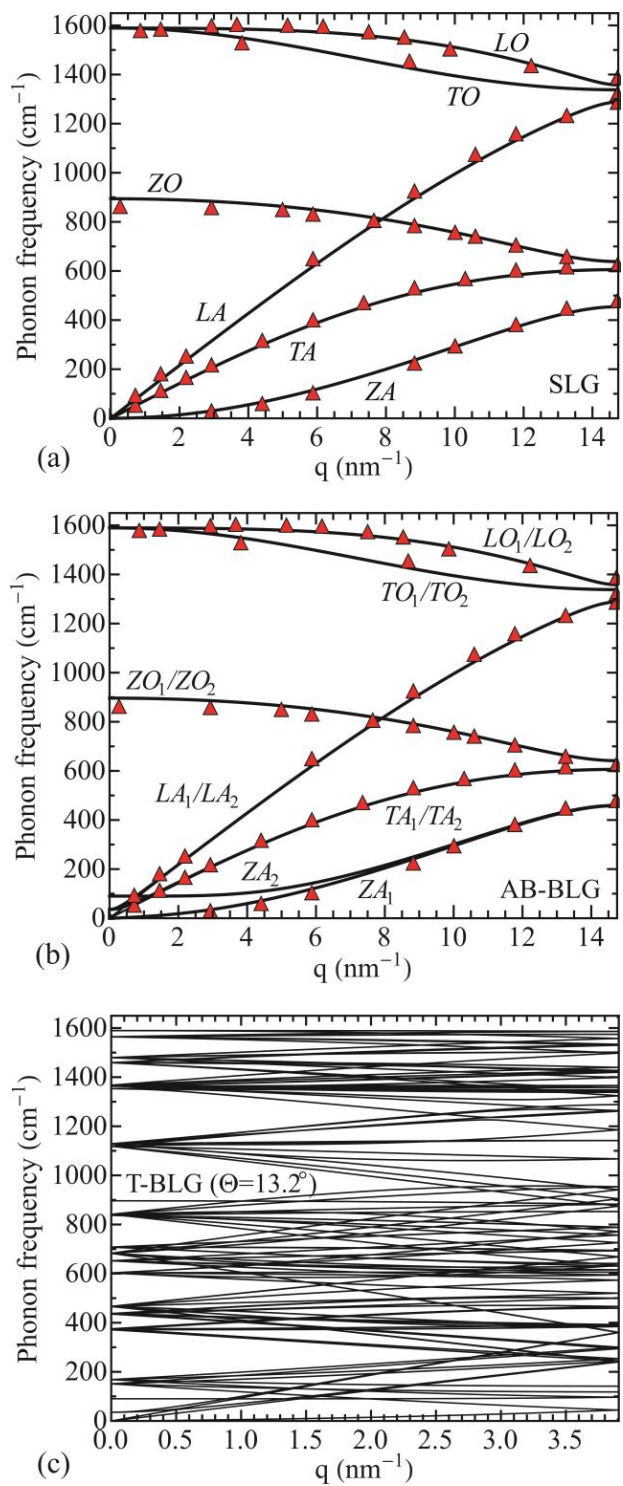

Figure 2 of 5: A.I. Cocemasov et al.

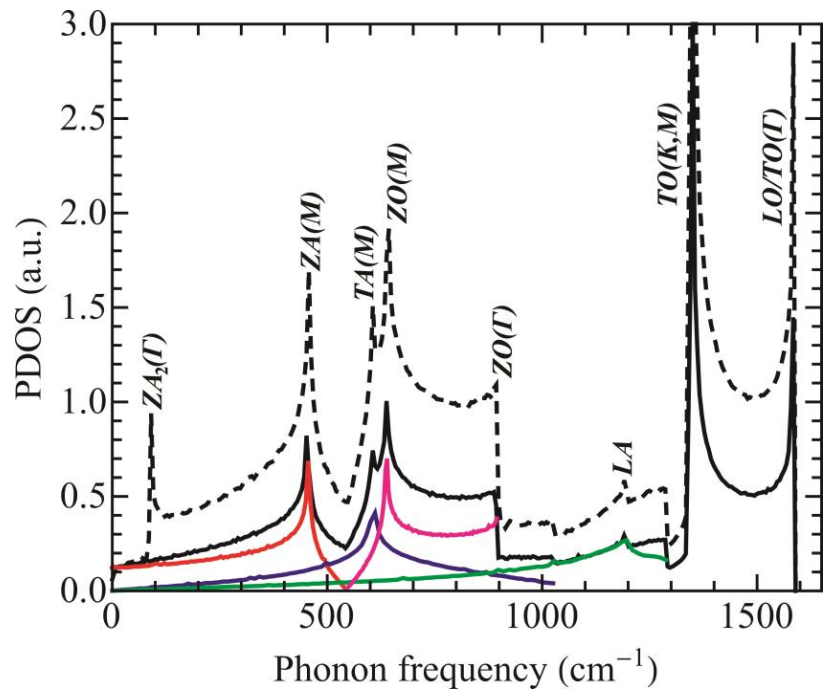

Figure 3 of 5: A.I. Cocemasov et al.

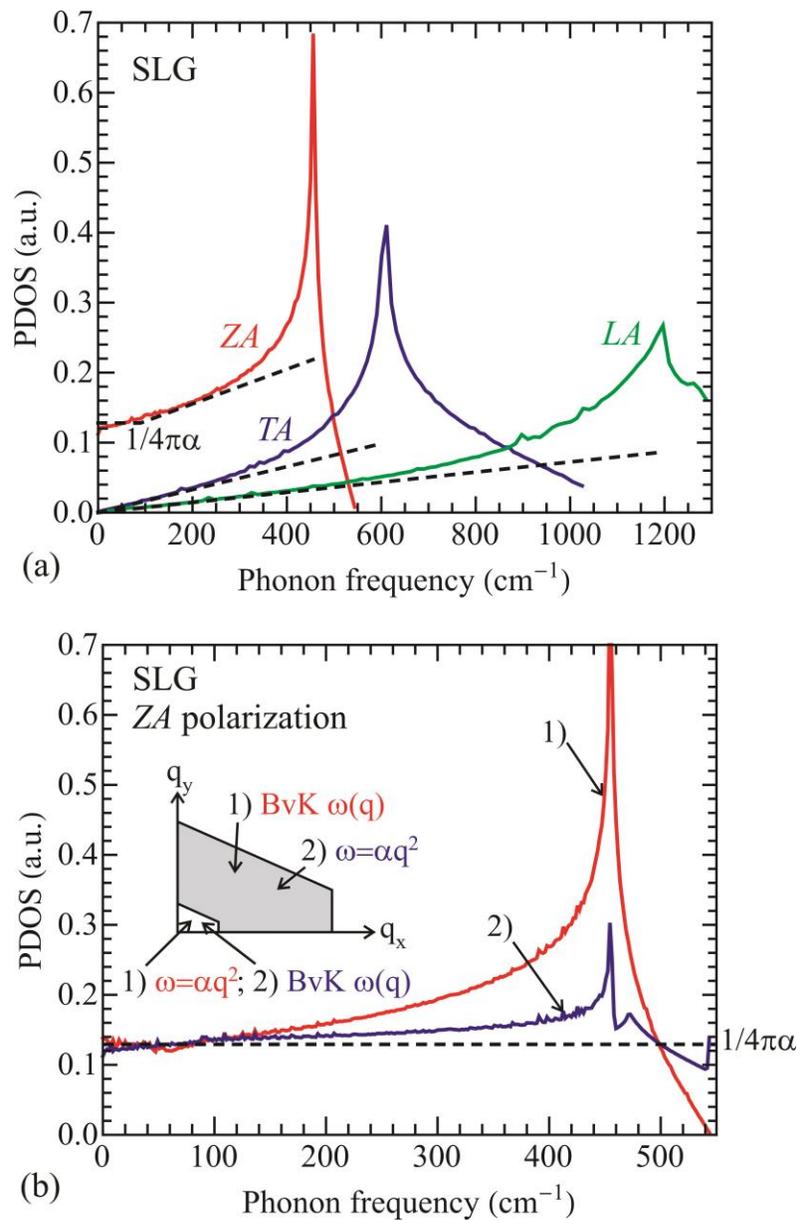

Figure 4 of 5: A.I. Cocemasov et al.

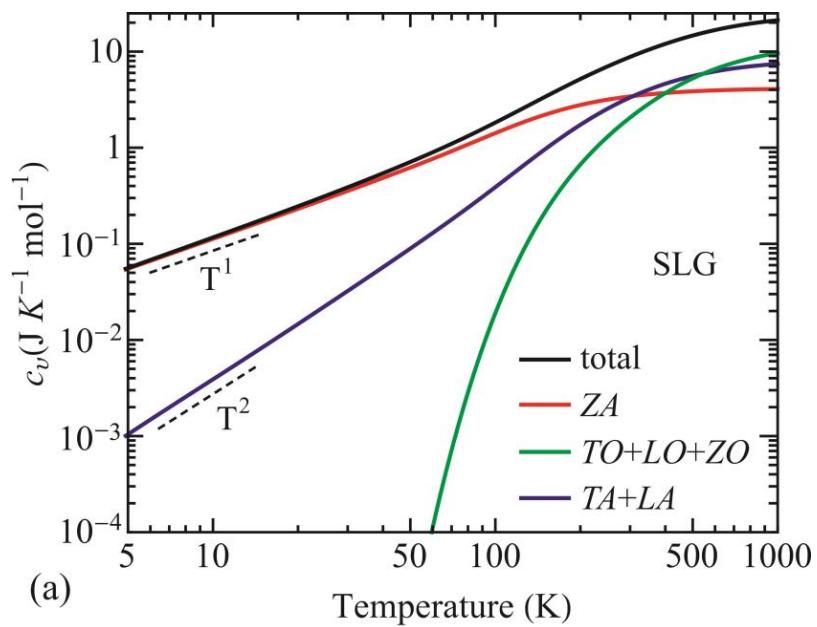

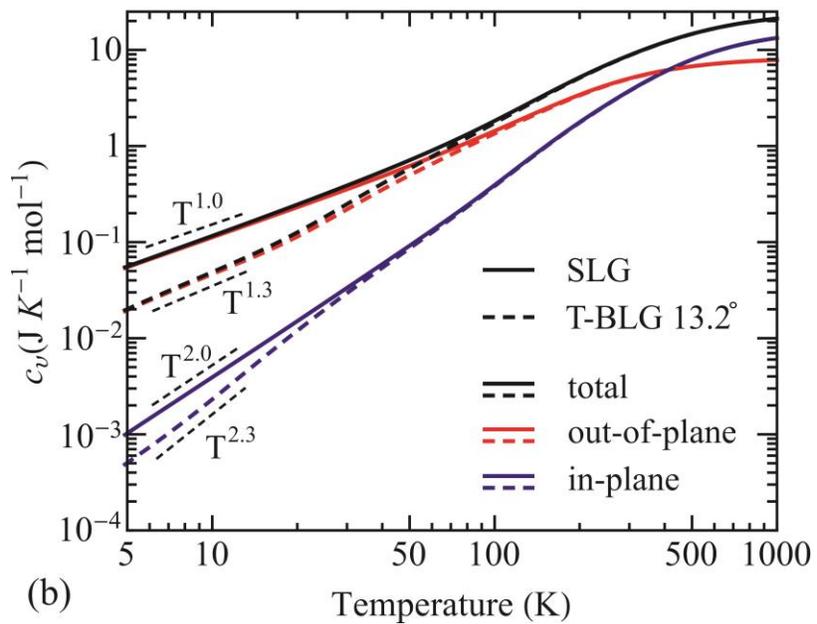

Figure 5 of 5: A.I. Cocemasov et al.